 \documentclass[final,5p,times,twocolumn]{elsarticle}

 \usepackage{graphics}
 \usepackage{graphicx}

\usepackage{amssymb}

\journal{Journal of Molecular Spectroscopy}

\begin{document}

\begin{frontmatter}



\title{Rotational spectroscopy, dipole moment and $^{14}$N nuclear
  hyperfine structure of $iso$-propyl cyanide\tnoteref{Widmung}}
\tnotetext[Widmung]{We dedicate this work to A. Robert W. McKellar, 
to Philip R. Bunker, and also to James K.G. Watson, who chose 
not to be officially named in this special issue, for their 
extensive contributions to experimental and theoretical 
molecular spectroscopy.}


\author[Koeln]{Holger S.P. M\"uller\corref{cor}}
\ead{hspm@ph1.uni-koeln.de}
\cortext[cor]{Corresponding author.}
\author[Toulouse1,Toulouse2]{Audrey Coutens}
\author[Toulouse1,Toulouse2]{Adam Walters}
\author[Hannover]{Jens-Uwe Grabow}
\author[Koeln]{Stephan Schlemmer}

\address[Koeln]{I.~Physikalisches Institut, Universit{\"a}t zu K{\"o}ln, 
   Z{\"u}lpicher Str. 77, 50937 K{\"o}ln, Germany}
\address[Toulouse1]{Universit\'e de Toulouse; UPS-OMP; IRAP;  Toulouse, France}
\address[Toulouse2]{CNRS; IRAP; 9 Av. colonel Roche, BP 44346, 31028 Toulouse cedex 4, France}
\address[Hannover]{Institut f\"ur Physikalische Chemie \& Elektrochemie, Lehrgebiet~A, 
Gottfried-Wilhelm-Leibniz-Universit\"at, Callinstr. 3A, 30167 Hannover, Germany.}

\begin{abstract}

Rotational transitions of $iso$-propyl cyanide, (CH$_3$)$_2$CHCN, also known as 
$iso$-butyronitrile, were recorded using long-path absorption spectroscopy 
in selected regions between 37 and 600~GHz. 
Further measurements were carried out between 6 and 20~GHz employing Fourier transform 
microwave (FTMW) spectroscopy on a pulsed molecular supersonic jet. The observed transitions 
reach $J$ and $K_a$ quantum numbers of 103 and 59, respectively, and yield accurate 
rotational constants as well as distortion parameters up to eighth order. The $^{14}$N 
nuclear hyperfine splitting was resolved in particular by FTMW spectroscopy yielding 
spin-rotation parameters as well as very accurate quadrupole coupling terms. In addition, 
Stark effect measurements were carried out in the microwave region to obtain a 
largely revised $c$-dipole moment component and to improve the $a$-component. 
The hyperfine coupling and dipole moment values are compared with values for related 
molecules both from experiment and from quantum chemical calculations.

\end{abstract}


\end{frontmatter}




\section{Introduction}
\label{introduction}

Molecules containing the cyanide group, -CN, are quite numerous in space. 
Among the approximately 160 molecules detected in the interstellar medium or in 
circumstellar shells of late-type stars there are almost 30 molecules containing 
a cyano group, see for example the Molecules in Space web page\footnote{Internet address: 
https://cdms.astro.uni-koeln.de/classic/molecules} of the Cologne Database for Molecular 
Spectroscopy, CDMS~\cite{CDMS_1,CDMS_2}. These include small inorganic species such 
as SiCN, cyanopolyynes HC$_{2n-1}$N up to HC$_{11}$N, as well as radicals and anions 
derived from the shorter cyanopolyynes. Saturated cyanides have been detected as well. 
Methyl cyanide, CH$_3$CN was among the earliest molecules to be detected in space by 
means of radio astronomy~\cite{MeCN-det}. Transitions of rarer isotopologs, 
including CH$_2$DCN~\cite{CH2DCN-det}, have also been recorded. 
The heavier ethyl cyanide, C$_2$H$_5$CN, was detected in space more than 30 years 
ago~\cite{EtCN-det}, and isotopologs containing $^{13}$C have been identified 
in Orion~KL~\cite{13C-EtCN-det_Orion-KL} and in Sagittarius~B2(N) (Sgr~B2(N) 
for short), where also corresponding isotopic species of vinyl cyanide were 
detected~\cite{13C-EtCN-det_Sgr_B2}. 
The latter two reports result from molecular line surveys of these two intensely 
studied high-mass star-forming regions carried out with the IRAM~30\,m radio telescope 
on the Pico del Veleta near Granada, Spain in order to investigate the molecular 
complexities of these two prolific sources.


\begin{figure}
 \begin{center}
  \includegraphics[width=7.0cm]{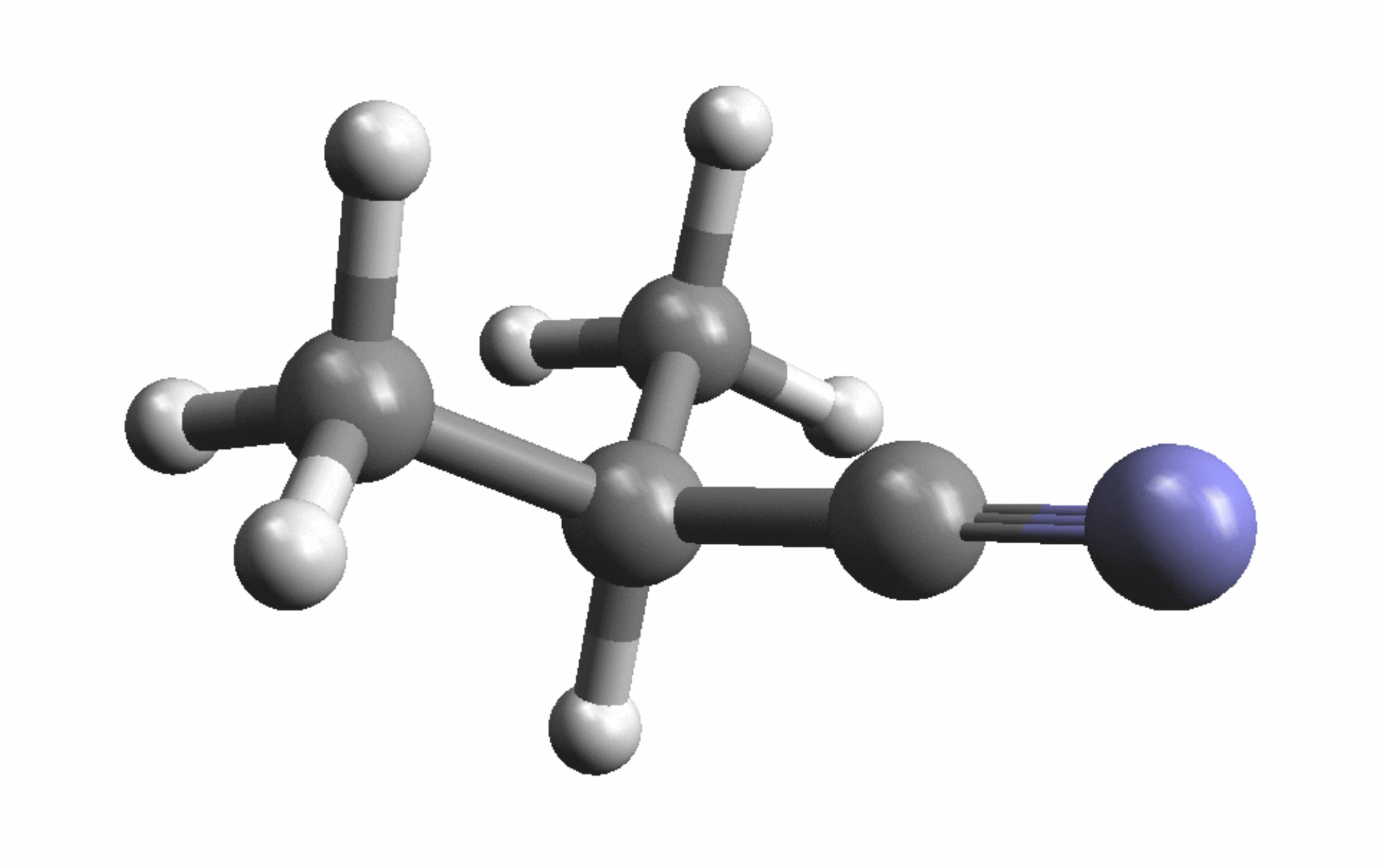}
 \end{center}
  \caption{Schematic depiction of the molecular structure of $iso$-propyl cyanide.}
  \label{structure}
\end{figure}

A part of the Sgr~B2(N) region is called Large Molecule Heimat because many of the 
larger molecules have been detected there for the first time. In fact, the molecular 
line survey mentioned above, has revealed to date three new large molecules. 
The first one, aminoacetonitrile~\cite{AAN-det}, is proposed to be the precursor for 
glycine, the simplest aminoacid, a molecule awaiting its detection by radio-astronomy. 
The two others ones are ethyl formate and $n$-propyl cyanide, 
$n$-C$_3$H$_7$CN~\cite{n-PrCN_EtFo-det}; the latter being the next larger member 
in the series of alkyl cyanides.

Propyl cyanide, also known as butyronitrile, is the first alkyl cyanide 
which may occur in more than one isomeric form. 
The cyano group is attached to the first carbon atom of the propyl group in the $n$- 
or $normal$-form, thus yielding an unbranched chain, whereas it is attached to the 
second C-atom in $iso$-propyl cyanide, yielding a branched structure, as shown in 
Fig.~\ref{structure}. 
After the detection of $n$-propyl cyanide in space it was logical to look into 
the ratio between the chain isomer and the branched isomer of propyl cyanide 
which should provide insight into the astrochemistry of complex molecules.

There had been only two published investigations into the rotational spectrum of 
$iso$-propyl cyanide. Herberich~\cite{i-PrCN_rot_1967} recorded more than 50 rotational, 
mostly $Q$-branch, transitions between 5 and 30~GHz with the $J$ quantum numbers 
up to 57, but published only 22 with $J \leq 14$ and $K_a \leq 5$. 
The $a$-type spectrum was expected to be about 20 times stronger than the $c$-type 
spectrum, i.e. $|\mu _a/\mu _c| \approx 4.5$. The observations seemed to be in accord 
with these expectations, and, consequently, most of the published lines were $a$-types. 
Uncertainties were not stated, but seem to be much better than 100~kHz, 
possibly as low as 20~kHz for some of them. Quite accurate $^{14}$N nuclear quadrupole 
coupling parameters were obtained from splitting observed in some transitions. 
No centrifugal distortion analysis was performed, but rotational constants were given.

Durig and Li~\cite{i-PrCN_rot_1974} presented 10 $R$-branch transition frequencies 
obtained between 28 and 38~GHz along with 8 transition frequencies each for 
the three lowest vibrational states. The accuracies were estimated to be 0.2~MHz. 
They performed some structural evaluations and determined the dipole moment components 
from Stark effect measurements of three $J = 3 - 2$ $a$-type transitions with 
$K_a \leq 1$ as $|\mu _a| = 4.05 \pm 0.02$~D and $|\mu _c| = 1.4 \pm 0.2$~D.
Neither work resolved possible splitting caused by internal rotation of the methyl groups.

A combined fit of these two data sets permitted some centrifugal distortion parameters 
to be determined, but the resulting predictions were not accurate enough to search 
for the molecule in the 3~mm region of the spectral survey of Sgr~B2(N), 
the lowest frequency region that was also deemed to be the best to search 
for such complex molecules. 
Therefore, an extensive laboratory study has been initiated 
to record rotational transitions in selected frequency windows from the microwave 
to the submillimeter region to obtain very accurate rotational and 
centrifugal distortion parameters as well as greatly improved $^{14}$N hyperfine 
(hf) coupling values. Noting that the $c$-type transitions appeared to be much weaker 
than predicted from the published dipole moment components, we also carried out 
Stark effect measurements which have resulted in large improvement for both components 
and yielded a much smaller value for the $c$-component.


\section{Experimental details}
\label{exp}

Absorption spectra of $iso$-propyl cyanide in the millimeter (mm) and submillimeter 
(sub-mm) regions were recorded at the Universit{\"a}t zu K{\"o}ln. The $37 - 69$~GHz 
range was covered with a commercial synthesizer (Agilent E8257D) as source, 
which nominally extends from almost 0 to 67~GHz. 
Its output was amplified and frequency-stabilized by a phase-lock loop (PLL) referenced 
to a rubidium reference. The lower frequency cut-off was caused by the amplifier. 
A Schottky-diode was used as detector. 
The amplified output of the same synthesizer was used to drive a commercial 
frequency quadrupler (VDI-AMC-S150e) and a superlattice frequency multiplier 
\cite{superlattice_2007} in series for investigations between 303 and 346~GHz. 
Additional measurements were performed in the frequency range 589$-$600~GHz 
with the Cologne Terahertz Spectrometer \cite{CTS1994} using a phase-locked 
backward-wave oscillator as source.
A liquid He-cooled InSb hot-electron bolometer was used as detector in the 
sub-mm region. To increase the sensitivity, frequency-modulation was employed 
throughout. Demodulation at twice the frequency causes a line-shape which is 
close to the second derivative of a Gaussian.

A 6~m long absorption cell was used in the lower two frequency regions while a
3~m long cell was employed near 600~GHz. A commercial sample of $iso$-propyl cyanide 
(Sigma-Aldrich) was used without further purification at pressures of around 0.1~Pa 
to around 1~Pa. All measurements were carried out at room temperature.


Spectra of $i$-C$_3$H$_7$CN between 6.8 and 19.9~GHz were recorded at the 
Gottfried-Wilhelm-Leibniz-Universit\"at in Hannover using a supersonic-jet 
Fourier transform microwave (FTMW) spectrometer \cite{Balle-Flygare-type} 
in the coaxially oriented beam-resonator arrangement (COBRA) \cite{COBRA-type} 
using a very broadband (2$-$26.5~GHz) and very sensitive set-up 
\cite{Multi-Octave-type,Critically-Coupled-type}. 
The sample was highly diluted with Ne to a total pressure of 100~kPa and expanded 
through a solenoid valve (General Valve series~9, nozzle orifice diameter 1.2~mm) 
into the Fabry-P{\'e}rot type resonator. 
A repetition rate of 20~Hz was used for the pulsed expansion. 
Following transient absorption, the spectral positions of the MW transient emission 
were determined after Fourier transformation of the 4k-data-point 
time-domain signal recorded at intervals of 100~ns, being 4k-zero-padded 
prior to transformation. Each molecular signal is split by the Doppler effect 
as a result of the coaxial arrangement of the supersonic jet and resonator axes 
in the COBRA set-up. 
The rest frequencies are calculated as the arithmetic mean of the frequencies 
of the two Doppler components. The accuracies range from 0.3 to 0.5~kHz because of 
the very high signal-to-noise ratio and very favorable, symmetric line shapes. 

Stark effect measurements were performed with the same spectrometer employing the 
coaxially aligned electrodes for Stark-effect applied in resonators (CAESAR) 
setup \cite{caesar_2004}, which provides a fairly homogeneous electric field 
over the entire volume, from which molecules are effectively contributing 
to the emission signal.


\section{Quantum chemical calculations}
\label{B3LYP}

The commercially available program Gaussian~03 \cite{Gaussian03} was used mainly 
to calculate the $^{14}$N nuclear quadrupole tensors for $iso$-propyl cyanide and 
two related molecules in their inertial and principal axes systems and 
also to compare other experimental spectroscopic parameters with quantum-chemically 
evaluated ones. The popular hybrid density-functional theory method B3LYP and 
the MP2 method were used. The correlation consistent basis set aug-cc-pVTZ 
of triple zeta quality was used for tight convergence structure calculations 
and for harmonic force field calculations which also yield equilibrium quartic centrifugal 
distortion parameters. As the program usually does not put molecules into the inertial 
axis system, the molecules were rotated accordingly, and dipole moments and hyperfine 
parameters were evaluated requesting no symmetry to be used. Core-correlating basis 
functions were included for the latter calculations to arrive at the aug-cc-pwCVTZ 
basis set \cite{core-correlating}. The MP2 density was requested for the corresponding 
calculations, as this was not the default. These calculations were performed with the 
core electrons frozen, which is the default. 
Trial calculations of the nuclear quadrupole properties at the MP3 level without structure 
optimization turned out to be barely justifiable in terms of required time and results. 
These calculations were done at the MP2/aug-cc-pVTZ structures. Corresponding calculations 
with structure optimization or even higher level calculations were deemed to be 
too time-consuming. Finally, the nuclear spin-rotation 
parameters (requesting output=Pickett) are provided defined negatively which is in 
accord with the NMR conventions and with the initial spectroscopic conventions, 
but nowadays, they are given almost exclusively defined positively in spectroscopy.


\begin{figure}
  \includegraphics[width=8.8cm]{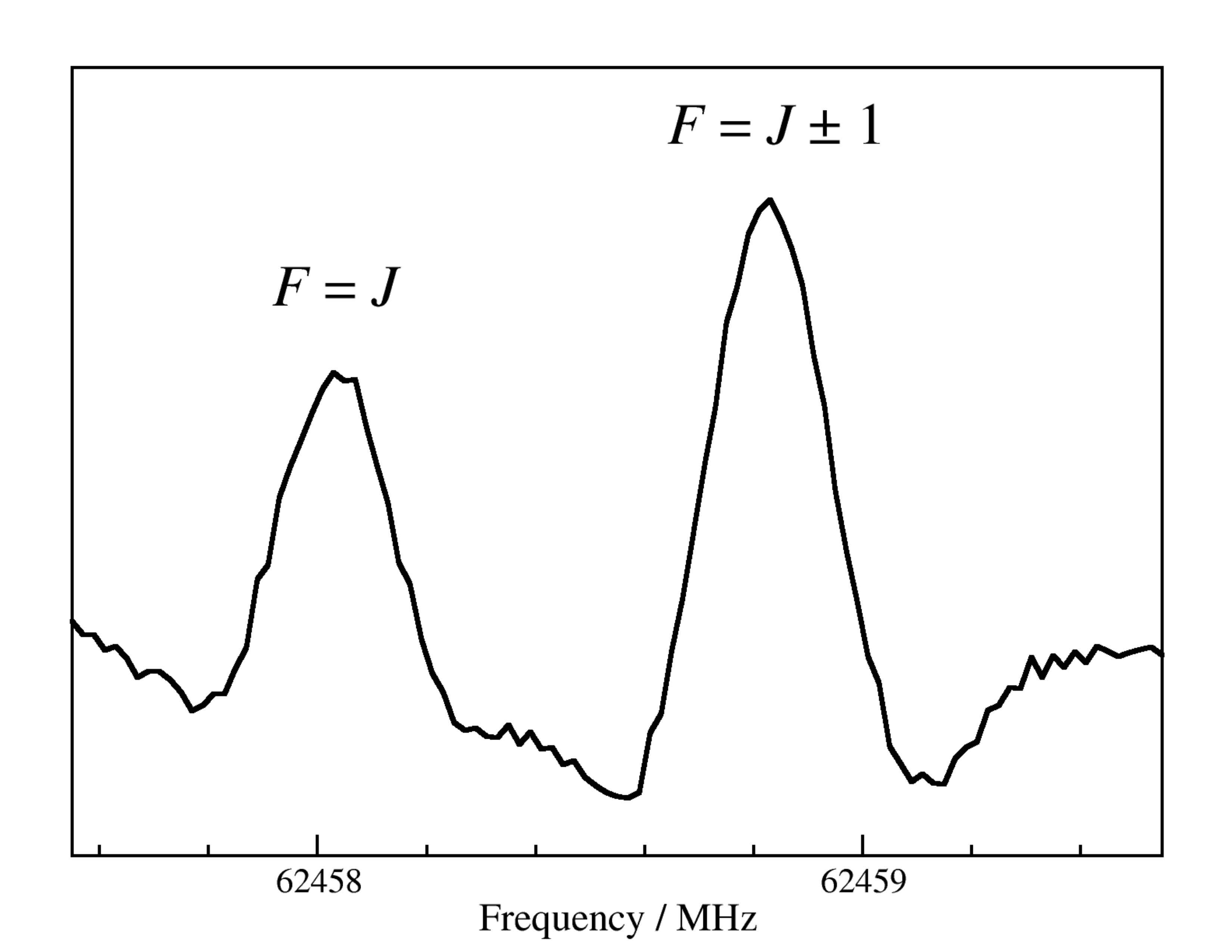}
  \caption{The prolate paired $J_{K_a} = 9_8 - 8_8$ transitions of $iso$-propyl cyanide 
    in the millimeter region showing partially resolved $^{14}$N hyperfine (hf) splitting; 
    the $F$ quantum numbers are given. Prolate paired transitions are those for which 
    the two transitions involving levels with $K_c = J + K_a$ and $K_c = J + K_a + 1$ 
    are not resolved; the $K_c$ quantum numbers are frequently omitted in these cases. 
    The role of $K_a$ and $K_c$ is reversed in oblate paired transitions.}
  \label{spectrum:mmW}
\end{figure}


\section{Observed spectrum and analysis}
\label{obs_spectrum}

\begin{figure}
  \includegraphics[width=8.8cm]{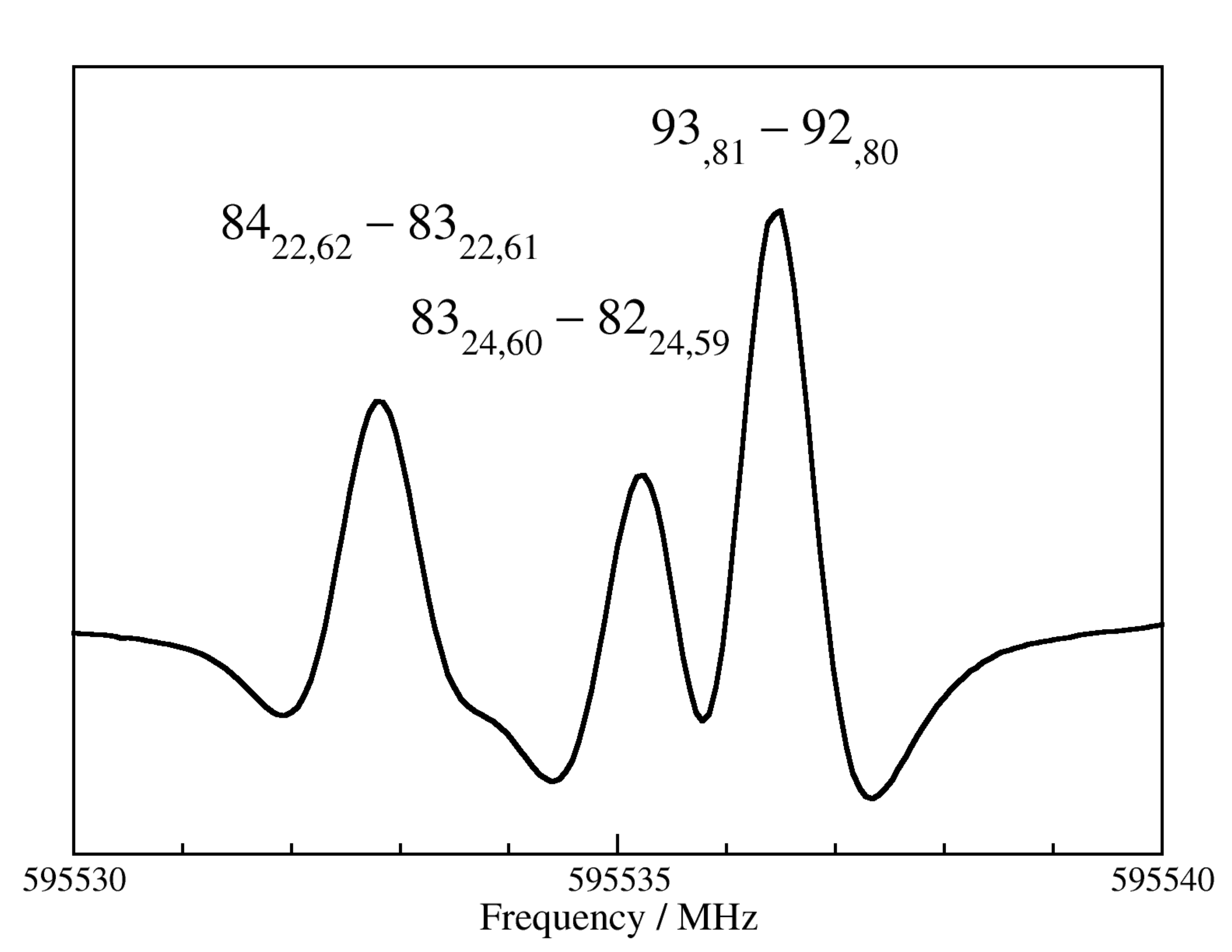}
  \caption{Detail of the sub-millimeter spectrum of $iso$-propyl cyanide showing 
   two individual and one oblate paired transitions with the $K_a$ quantum 
   numbers for the latter being omitted; see also caption to Fig.~\ref{spectrum:mmW}.}
  \label{spectrum:sub-mmW}
\end{figure}

The $iso$-propyl cyanide molecule is a rather asymmetric rotor of the prolate type 
($\kappa = -0.5766$) with a large $a$-dipole moment component and a still sizable 
$c$-component as mentioned in section~\ref{introduction}. 
Previous studies~\cite{i-PrCN_rot_1967,i-PrCN_rot_1974} did not resolve any 
methyl internal rotation splitting. In fact, even the resolution attainable with 
FTMW spectroscopy (line widths of order of 3~kHz) did not reveal any such splitting. 
The spin of the $^{14}$N nucleus is 1, leading to nuclear quadrupole splitting, 
which at lower frequencies and high resolution can be resolved. The magnetic 
moment of this nucleus can modify the quadrupole splitting because of the 
nuclear spin-rotation coupling. However, for a molecule with such small rotational 
constants and for a nucleus with such small magnetic moment, such effects are expected 
to be very small. $I(^{14}{\rm N}) = 1$ causes each rotational level to be split into 
three hf levels. The hf selection rules are $\Delta F = 0, \pm1$. The strong components 
have $\Delta F = \Delta J$, but the weaker components can be observed under favorable 
conditions, in particular for low-$J$ transitions. At somewhat higher quantum numbers, 
the $F = J \pm 1$ components may overlap while the one with $F = J$ may still be 
resolvable as shown in Fig.~\ref{spectrum:mmW}. Eventually no hf splitting can 
be resolved (e.g. Fig.~\ref{spectrum:sub-mmW}) unless one resorts to sub-Doppler techniques.


Predictions from a combined fit of previous data were good enough to search for some 
lower-$J$ quantum number $a$-type $R$-branch transitions in small frequency windows 
of the lower mm region ($37 - 69$~GHz). 
Pickett's {\scriptsize SPCAT} and {\scriptsize SPFIT} programs \cite{Herb} 
were used for prediction and fitting of the spectra, respectively.
With refined predictions, transitions with higher $J$ and $K_a$ were recorded. 
Later, $a$-type $Q$-branch transitions, first with $\Delta K_a = 0$, 
subsequently with $\Delta K_a = 2$, were found together with very few unexpectedly 
weak $c$-type transitions. 
Even though the molecule has several low-lying vibrational modes, the spectrum was 
sparse enough in the lower millimeter region that overlap of transitions searched for 
occured only on few occasions and mostly for rather weak transitions. 
Some of the transitions showed splitting caused by the $^{14}$N nuclear hyperfine 
coupling as can be seen in Fig.~\ref{spectrum:mmW}. Unresolved asymmetry splitting 
occured frequently for transitions having relatively high $K_a$ and low $J$ values 
(prolate pairing, shown in Fig.~\ref{spectrum:mmW}) or, at higher frequencies, 
for those with low $K_a$ and high $J$ values (oblate pairing, 
shown in Fig.~\ref{spectrum:sub-mmW}).

Subsequently, measurements of selected transitions were performed in the lower 
($303 - 346$~GHz) and somewhat higher sub-mm region ($589 - 600$~GHz). 
These were mostly $a$-type transitions, see Fig.~\ref{spectrum:sub-mmW}, but also 
a considerable number of $c$-type transitions. Transitions having $\Delta K_a \geq 2$ 
were not searched for, except in one case, because they were rather weak, and the 
predicted uncertainties were usually smaller than those of $c$-type transitions of 
similar strengths. Quantum numbers $J$ and $K_a$ as high as 103 and 59, respectively, 
have been accessed.


\begin{figure}
  \includegraphics[width=8.8cm]{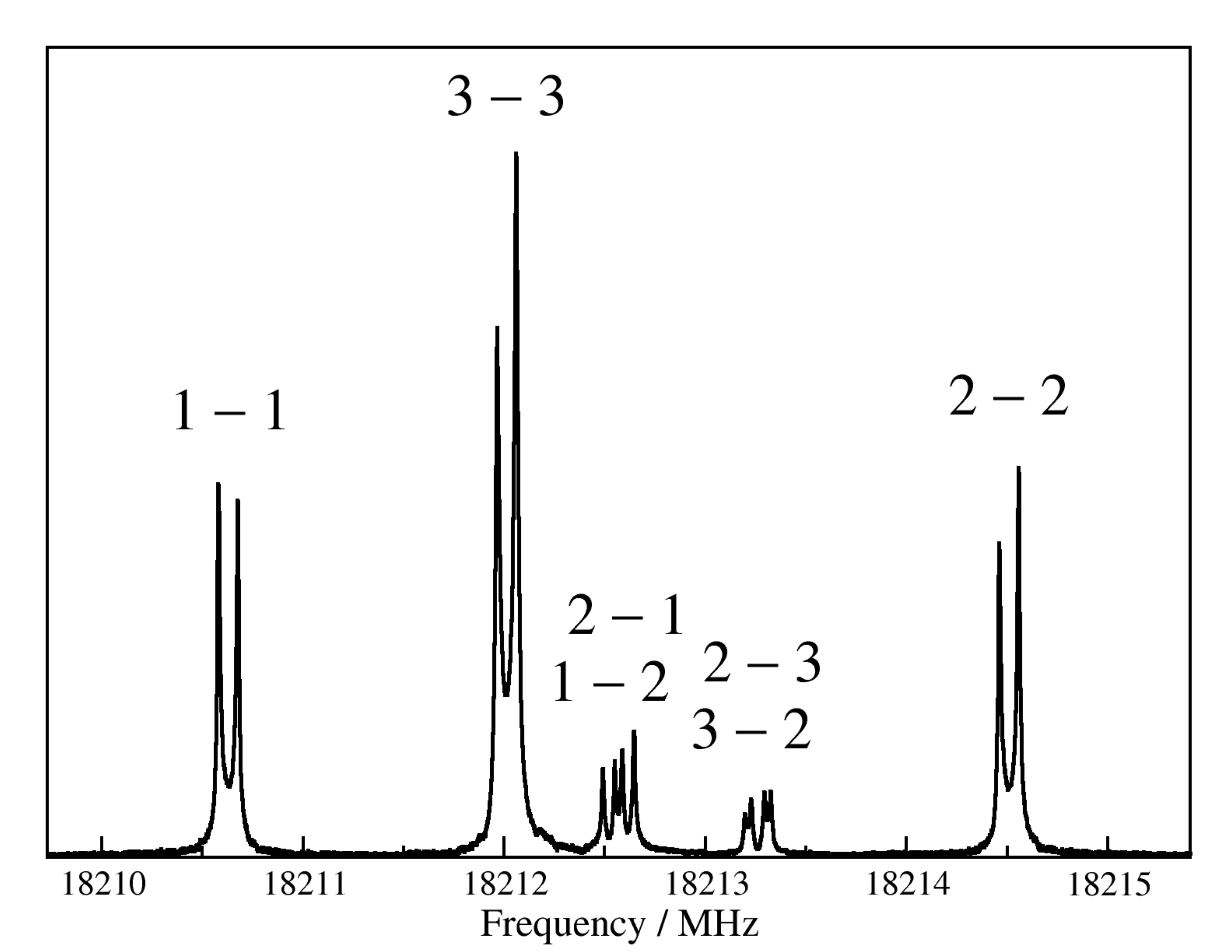}
  \caption{The $J_{K_a,Kc} = 2_{2,1} - 2_{0,2}$ transition of $iso$-propyl cyanide 
   recorded in several sections by molecular beam Fourier transform microwave spectroscopy. 
   The $^{14}$N  hf splitting is well resolved; the $F' - F''$ quantum numbers are given. 
   Each line is split into two Doppler components because the microwave radiation 
   propagates in the same direction as the supersonic jet. 
   Because of the very high sensitivity, not only the three strong $\Delta F = 0$, 
   but also the four weaker $\Delta F \neq 0$ hf components are seen.}
  \label{spectrum:FTMW}
\end{figure}

The observation of the $c$-type transitions being much weaker than predicted from the 
published dipole moment components prompted new Stark effect measurements by FTMW 
spectroscopy as described in section~\ref{Dipole}. We used the opportunity 
to record some transitions in the absence of an electric field in order 
to improve the hyperfine parameters as well as the lower order spectroscopic parameters. 
The very high sensitivity of the spectrometer permitted to observe not only the strong 
$\Delta F = \Delta J$ hf components, but also the weaker ones, even for transitions 
already not so strong as is demonstrated in Fig.~\ref{spectrum:FTMW}.

468 transitions belonging to slightly more than 300 distinct spectral features were 
used in the final fit. 41 hf components of 9 rotational transitions were obtained 
in the microwave region. Of the more than 100 transitions from the millimeter region 
almost half represent partially resolved hf features, some of which showed prolate 
pairing. More than 150 transitions each were recorded slightly above 300 and below 
600~GHz, respectively; several of these are unresolved asymmetry doublets. 
The few previously reported transition frequencies were omitted from the final fit 
as their uncertainties were rather large in one case and rather uncertain in the other. 
Moreover, trial fits showed that omission of these data had a negligible effect on 
the values as well as the uncertainties of the spectroscopic parameters.


\begin{table}
  \caption{Experimentally determined spectroscopic parameters$^a$ (MHz) of 
           $iso$-propyl cyanide in comparison to those from quantum chemical 
           calculations$^b$}
  \label{parameters}
{\footnotesize
  \begin{tabular}{lr@{}lr@{}lr@{}l}
  \hline
   Parameter & \multicolumn{2}{c}{Experimental value} & \multicolumn{2}{c}{B3LYP} 
   & \multicolumn{2}{c}{MP2} \\
  \hline
$A$                      &   7940&.877174~(31)    &   7934&.0   &   8022&.5   \\
$B$                      &   3968&.087775~(27)    &   3974&.1   &   3936&.3   \\
$C$                      &   2901&.053223~(22)    &   2899&.0   &   2914&.9   \\
$D_K \times 10^3$        &   $-$5&.23242~(61)     &   $-$4&.50  &   $-$5&.50  \\
$D_{JK} \times 10^3$     &     12&.17725~(42)     &     11&.55  &     12&.62  \\
$D_J \times 10^6$        &    610&.2684~(153)     &    593&.    &    610&.    \\
$d_1 \times 10^6$        & $-$244&.0908~(69)      & $-$240&.    & $-$239&.    \\
$d_2 \times 10^6$        & $-$189&.2889~(76)      & $-$180&.    & $-$189&.    \\
$H_K \times 10^9$        &  $-$37&.163~(248)      &       &     &       &     \\
$H_{KJ} \times 10^9$     &     10&.968~(281)      &       &     &       &     \\
$H_{JK} \times 10^9$     &     38&.267~(140)      &       &     &       &     \\
$H_J \times 10^{12}$     & $-$584&.83~(151)       &       &     &       &     \\
$h_1 \times 10^{12}$     &     68&.73~(77)        &       &     &       &     \\
$h_2 \times 10^{12}$     &    884&.23~(83)        &       &     &       &     \\
$h_3 \times 10^{12}$     &    322&.39~(140)       &       &     &       &     \\
$L_{KKJ} \times 10^{15}$ & $-$242&.~(61)          &       &     &       &     \\
$L_{JJK} \times 10^{15}$ &  $-$79&.7~(95)         &       &     &       &     \\
$l_3 \times 10^{15}$     &   $-$1&.658~(148)      &       &     &       &     \\
$l_4 \times 10^{15}$     &   $-$0&.607~(41)       &       &     &       &     \\
$\chi _{aa}$             &   $-$3&.93838~(23)     &   $-$4&.380 &   $-$3&.584 \\
$\chi _{bb}$             &      2&.11117~(30)     &      2&.338 &      1&.919 \\
$\chi _{cc}$             &      1&.82721~(24)$^c$ &      2&.042 &      1&.665 \\
$C_{aa} \times 10^3$     &      0&.162~(76)       &      0&.2   &      0&.1   \\
$C_{bb} \times 10^3$     &      0&.834~(97)       &      0&.9   &      0&.8   \\
$C_{cc} \times 10^3$     &      0&.583~(74)       &      0&.6   &      0&.5   \\
    \hline
  \end{tabular}\\[2pt]
}
$^a$\footnotesize{Numbers in parentheses are one 
  standard deviation in units of the least significant figures.}\\
$^b$\footnotesize{See section~\ref{B3LYP} for details.}\\
$^c$\footnotesize{Derived parameter.}\\
\end{table}

Rotational constants along with complete sets of quartic and sextic as well as some 
octic centrifugal distortion parameters were determined employing Watson's $S$-reduction 
of the rotational Hamiltonian. The $A$-reduction yielded a slightly worse result 
even with one parameter more in the fit. The highly accurate data obtained by 
FTMW spectroscopy not only permitted $^{14}$N nuclear quadrupole coupling values 
to be evaluated, but also yielded nuclear spin-rotation coupling constants. 
Trial fits with the off-diagonal quadrupole term $\chi _{ac}$ yielded 
$-1.06 \pm 1.57$ for its magnitude; the sign cannot be determined from the fit, 
but can often be deduced from the values of the diagonal terms. 
Even though this value appears to be reasonable, the parameter was omitted 
from the final fit as it was not even close to being determined with significance. 
In addition, inclusion of the parameter did have a negligible effect 
on the quality of the fit and on the values of all other parameters. 
The final spectroscopic parameters are given in Table~\ref{parameters} 
together with values from quantum chemical calculations as far as available. 
The rms error of the fit is 0.85 and differs little for subsets of the data 
(between 0.74 and 0.96), indicating essentially appropriate estimates 
of the uncertainties. The measured transition frequencies with their assignments, 
uncertainties, and residuals between observed values and those calculated from 
the final set of spectroscopic parameters are given in the supplementary 
material. Predictions of the rotational spectrum along with a documentation will 
be provided in the catalog section\footnote{Internet address: 
https://cdms.astro.uni-koeln.de/classic/entries/} of the Cologne Database for Molecular 
Spectroscopy, CDMS~\cite{CDMS_1,CDMS_2}. Line, parameter, and fit files, along with 
further auxiliary files, will be available in the data section\footnote{Internet address: 
https://cdms.astro.uni-koeln.de/classic/predictions/daten/i-PrCN/} of the CDMS, 
and can also be accessed via the archive part of the catalog.


\begin{figure}
  \includegraphics[width=8.8cm]{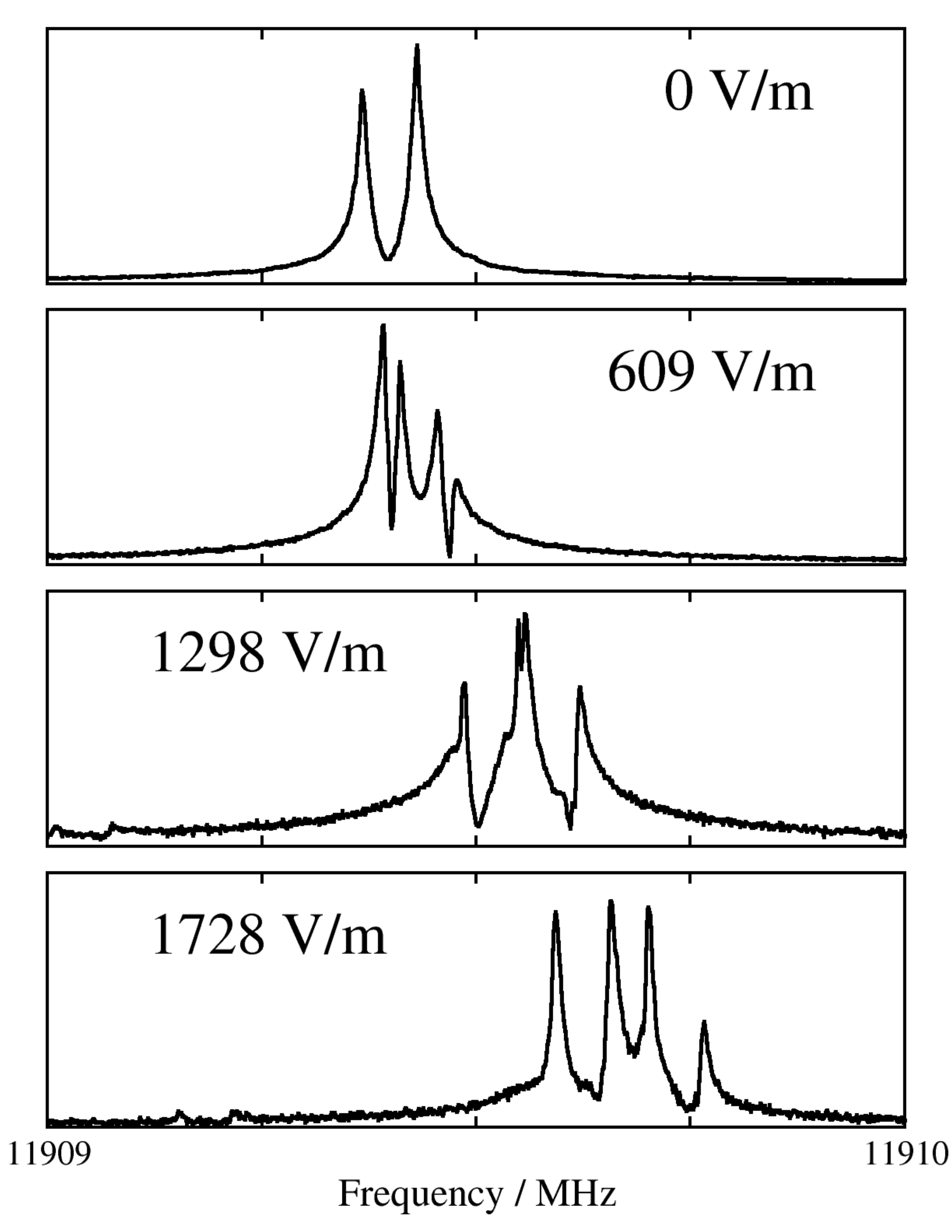}
  \caption{Stark effect measurements of the $F = 1 - 1$ hyperfine component pertaining 
   to the $J_{K_a,K_c} = 1_{1,0} - 0_{0,0}$ transition of $iso$-propyl cyanide. 
   Each molecular signal is split into two because of the Doppler effect; 
   see also caption to Fig.~\ref{spectrum:FTMW}. 
   The hf component splits into two $m$-components upon application of a 
   static electric field because of the $\Delta M = \pm1$ selection rules. 
   The electric field increases almost linearly from top to bottom, 
   and the frequency shifts display the second order or quadratic Stark effect.}
  \label{Stark_shifts}
\end{figure}


\section{Stark effect measurements}
\label{Dipole}

The Stark field is perpendicular to the polarization of the microwave field 
leading to the selection rules $\Delta M = \pm1$ \cite{caesar_2004}.
Transitions with small $J$ quantum numbers are frequently used for determining 
dipole moment components as these usually display large quadratic Stark shifts. 
In the present case, with hf splitting caused by the $^{14}$N nucleus, the 
$F' - F'' = 1 - 0$ (or $0 - 1$) and the $1 - 1$ components were deemed to be 
particularly suited for Stark effect measurements as these would lead to only 
one and two Stark components, respectively, thus minimizing the danger of line overlap. 
Fig.~\ref{Stark_shifts} demonstrates the second order Stark effect for one 
$1 - 1$ hf component which splits into two Stark components. Table~\ref{Stark}
summarizes the chosen transitions, hf, and Stark components as well as results 
from the analysis of the Stark effect. Other hf components were less suitable  
because of overlap already in the absence of an electric field or because overlap 
occured at rather low fields, both making frequency determinations less reliable.

The voltage readings were corrected marginally as described in ref.~\cite{Aminophenol_2008}. 
The distance calibration was also taken from that work; however, it was noticed that 
the dipole moment used for OC$^{36}$S was actually that of the main isotopolog involving 
$^{32}$S. The dipole moment of OC$^{36}$S has, to our knowledge, not been determined 
experimentally, but its value can be evaluated by applying the OC$^{34}$S/OC$^{32}$S 
dipole ratio to OC$^{34}$S. We used the case~(1) model data from 
ref.~\cite{OCS-dip_Tanaka2} to derive a value of 0.71564~D with an estimated 
uncertainty of about 0.05~mD.

The analysis was carried out with the program {\scriptsize QSTARK} \cite{qstark}.
Most of the Stark-shifted frequencies were given unit weight, some, associated with 
lines displaying higher noise or less favorable lineshapes were given slightly lower weights. 
Measurements with two Stark components very close to each other were not used 
in the fit as the frequencies were rather uncertain. Only the two non-zero dipole moment 
components $\mu _a$ and $\mu _c$ were allowed to vary in the fit. 
The statistical error was only 1.6~mD for $\mu _a$, a relative uncertainty of 
$4 \times 10^{-4}$. Thus, other sources of error should be considered. 
The limitations of determining the accuracies of the voltages should be included 
in the statistical uncertainties as very many Stark-shifts have been measured, 
thus, they should behave like statistical errors. 
The uncertainty in the constant voltage off-set should also matter little because 
it is not linearly dependent on the voltage. The neglect of the isotope dependence 
of the OCS dipole moment would have created a non-negligible error of 
$6.3 \times 10^{-4}$, but the individual OCS dipole moments have been determined 
to better than $10^{-4}$. Also, the uncertainty in the linear voltage correction is 
comparatively small ($1.5 \times 10^{-4}$). However, the two plate separations have 
uncertainties of about 5 and $9 \times 10^{-4}$, respectively, which both exceed 
the purely statistical error. The final uncertainty of $\mu _a$ has been increased 
to 5.0~mD, which is slightly larger than the maximum error of the plate separation 
combined with the statistical error.
Table~\ref{Stark} summarizes the Stark effect measurements while 
Table~\ref{Dipole_values} displays the experimental and theoretical dipole moment values 
for $iso$-propyl cyanide from this study together with values for two related molecules.


\begin{table}
  \caption{Stark effect measurement of $iso$-propyl cyanide giving
           first the quantum numbers of the rotational transition, 
           the hyperfine structure component, and the Stark component, 
           and subsequently the applied electric fields (V/m), 
           the measured frequencies (MHz), the weights of the measurement 
           in the fit, and the residuals O$-$C (kHz)}
  \label{Stark}
{\footnotesize
  \begin{tabular}{r@{}lr@{}lr@{}lr@{}l}
  \hline
\multicolumn{8}{l}{$J'_{K'_aK'_c} - J''_{K''_aK''_c}$, $F' - F''$, $M'_F - M''_F$} \\
  \hline
\multicolumn{2}{c}{Field} & \multicolumn{2}{c}{Frequency} &
\multicolumn{2}{c}{Weight} & \multicolumn{2}{c}{O$-$C}  \\
  \hline
\multicolumn{8}{l}{$1_{01} - 0_{00}$, $1 - 1$, $0 - 1$} \\
  \hline
     0&.0  &   6868&.15324  &   1&.00  & $-$0&.12       \\
  1280&.5  &   6868&.17406  &   0&.25  & $-$1&.36       \\
  1705&.0  &   6868&.19216  &   0&.25  & $-$0&.30       \\
  2299&.5  &   6868&.22285  &   0&.25  & $-$1&.63       \\[1mm]
\multicolumn{8}{l}{$1_{01} - 0_{00}$, $1 - 1$, $1 - 0$} \\
  \hline
     0&.0  &   6868&.15324  &   1&.00  & $-$0&.12       \\
  1280&.5  &   6868&.18955  &   0&.25  & $-$0&.80       \\
  1705&.0  &   6868&.21706  &   0&.25  & $-$1&.61       \\
  2299&.5  &   6868&.27052  &   0&.25  & $-$0&.72       \\[1mm]
\multicolumn{8}{l}{$1_{01} - 0_{00}$, $0 - 1$, $0 - 1$} \\
  \hline
     0&.0  &   6871&.10657  &   1&.00  &    0&.14       \\
   431&.5  &   6871&.11008  &   1&.00  &    0&.00       \\
   601&.0  &   6871&.11354  &   1&.00  &    0&.03       \\
   856&.0  &   6871&.12041  &   1&.00  & $-$0&.40       \\
  1280&.5  &   6871&.13795  &   0&.44  & $-$0&.72       \\
  1705&.5  &   6871&.16281  &   0&.25  & $-$0&.98       \\[1mm]
\multicolumn{8}{l}{$1_{10} - 0_{00}$, $0 - 1$, $0 - 1$} \\
  \hline
     0&.0  &  11908&.02845  &   1&.00  &    0&.54       \\
   437&.0  &  11908&.04335  &   1&.00  &    0&.35       \\
   609&.0  &  11908&.05792  &   1&.00  &    0&.87       \\
   867&.5  &  11908&.08663  &   1&.00  &    0&.35       \\
  1298&.0  &  11908&.15459  &   1&.00  &    0&.36       \\
  1728&.5  &  11908&.23989  &   1&.00  & $-$0&.70       \\[1mm]
\multicolumn{8}{l}{$1_{10} - 0_{00}$, $1 - 1$, $1 - 0$} \\
  \hline
     0&.0  &  11909&.39865  &   1&.00  &    0&.03       \\
   867&.5  &  11909&.44876  &   1&.00  &    0&.09       \\
  1298&.0  &  11909&.51770  &   1&.00  &    0&.51       \\
  1728&.5  &  11909&.62373  &   1&.00  &    0&.02       \\[1mm]
\multicolumn{8}{l}{$1_{10} - 0_{00}$, $1 - 1$, $0 - 1$} \\
  \hline
     0&.0  &  11909&.39865  &   1&.00  &    0&.03       \\
   867&.5  &  11909&.48263  &   1&.00  & $-$0&.05       \\
  1298&.0  &  11909&.58742  &   1&.00  &    0&.63       \\
  1728&.5  &  11909&.73230  &   1&.00  &    0&.03       \\[1mm]
\multicolumn{8}{l}{$2_{12} - 1_{11}$, $1 - 0$, $1 - 0$} \\
  \hline
     0&.0  &  12672&.69853   &  1&.00  & $-$0&.18       \\
   434&.0  &  12672&.71000   &  1&.00  &    0&.17       \\
   605&.5  &  12672&.72030   &  1&.00  & $-$0&.19       \\
   862&.0  &  12672&.74265   &  1&.00  & $-$0&.77       \\
  1289&.5  &  12672&.80158   &  1&.00  & $-$0&.15       \\
  1717&.5  &  12672&.88826   &  1&.00  &    0&.07       \\[1mm]
\multicolumn{8}{l}{$2_{11} - 1_{10}$, $1 - 0$, $1 - 0$} \\
  \hline
     0&.0  &  14806&.68033  &   1&.00  & $-$0&.60       \\
   435&.5  &  14806&.67043  &   1&.00  & $-$0&.91       \\
   607&.0  &  14806&.66176  &   1&.00  & $-$0&.71       \\
   864&.5  &  14806&.64461  &   1&.00  &    0&.36       \\
  1293&.0  &  14806&.60375  &   1&.00  &    0&.57       \\
  1722&.0  &  14806&.55352  &   1&.00  & $-$0&.73       \\[1mm]
\multicolumn{8}{l}{$2_{11} - 1_{01}$, $1 - 0$, $1 - 0$} \\
  \hline
     0&.0  &  19843&.60228  &   1&.00  & $-$0&.13       \\
   434&.0  &  19843&.60374  &   1&.00  & $-$0&.34       \\
   604&.5  &  19843&.60560  &   1&.00  & $-$0&.05       \\
   861&.5  &  19843&.60905  &   1&.00  &    0&.06       \\
  1289&.0  &  19843&.61691  &   1&.00  & $-$0&.19       \\
  1716&.5  &  19843&.62793  &   1&.00  & $-$0&.44       \\
  2315&.0  &  19843&.64930  &   1&.00  &    0&.00       \\
  2742&.5  &  19843&.66830  &   1&.00  &    0&.52       \\
  3255&.0  &  19843&.69289  &   1&.00  & $-$0&.72       \\
    \hline
  \end{tabular}
}
\end{table}


\section{Discussion}
\label{Discussion}

Accurate spectroscopic parameters have been determined for $iso$-propyl cyanide 
in the course of the present investigation. The rotational constants are consistent 
with the less accurate ones from previous studies \cite{i-PrCN_rot_1967,i-PrCN_rot_1974}, 
in particular taking into account that these studies did not perform centrifugal 
distortion analyses. As can be seen in Table~\ref{parameters}, the experimental rotational 
and quartic centrifugal distortion parameters agree favorably with the theoretical values, 
in particular if one considers that, first, the latter are equilibrium values, 
while the former are ground-state values, and second, the models are of limited sophistication. 
The very good agreement between the experimental and the B3LYP rotational constants 
compared to those obtained at the MP2 level may be fortuitous, though.
In the case of a molecule close to the prolate limit, $D_K$ is usually positive and 
shows a fairly strong variation with vibrational excitation, as can be seen for the 
chlorine dioxide molecule \cite{OClO_rot_1997}. In the present case, $D_K$ is negative 
and smaller in magnitude than $D_{JK}$. The somewhat larger deviation between experimental 
and B3LYP $D_K$ values of $iso$-propyl cyanide, compared with the remaining 
quartics, may well be explained by unaccounted vibrational effects. The very good 
agreement between some of the experimental and MP2 quartics is probably fortuitous.

The previously determined $^{14}$N nuclear quadrupole coupling parameters \cite{i-PrCN_rot_1967} 
are in quite good accord with the present, more accurate ones in Table~\ref{parameters}. 
The agreement with quantum-chemically calculated ones, also in that table, are reasonable. 
Again, the deviations may be due to vibrational effects and to short-comings 
of the theoretical methods. 
The magnitudes of the MP2 values are smaller than the experimental values 
by about the same amount as the B3LYP magnitudes are too large.
Vibrational corrections to nuclear quadrupole coupling values are often very small, 
as in HC$_3$N and DC$_3$N \cite{DC3N_rot_2008}, to moderate (a few percent), as in 
DCO$^+$ \cite{DCO+_DNC_astro-hfs}, H$_2 ^{17}$O \cite{H2O-17_hfs_2009}, and D$_2$O 
\cite{D2O_hfs_2010}. Thus, one would attribute the deviations between theoretical 
and experimental values of around 10\,\% more to deficiencies of the 
quantum chemical methods than to vibrational effects, at least in one of the two methods. 
It should be noted, though, that large vibrational effects ($\sim15\,\%$) are present 
in the $^{14}$N quadrupole coupling constant of HN$^{13}$C \cite{DCO+_DNC_astro-hfs} 
caused mainly by the floppy bending mode.


\begin{table*}
  \caption{Experimental dipole moment components (D) of $iso$-propyl cyanide in 
   comparison to values from quantum-chemical calculation and corresponding values 
   for the $anti$-conformer of $normal$-propyl cyanide and for $cyclo$-propyl cyanide$^a$}
  \label{Dipole_values}
{\footnotesize
  \begin{tabular}{lr@{}lr@{}lr@{}llr@{}lr@{}lr@{}llr@{}lr@{}lr@{}l}
  \hline
Component & \multicolumn{6}{c}{$i$-Propyl cyanide} & & \multicolumn{6}{c}{$a$-$n$-Propyl 
  cyanide} & & \multicolumn{6}{c}{$c$-Propyl cyanide} \\
\cline{2-7} \cline{9-14} \cline{16-21} 
 & \multicolumn{2}{c}{Exptl.} & \multicolumn{2}{c}{B3LYP} & \multicolumn{2}{c}{MP2} & &
\multicolumn{2}{c}{Exptl.$^b$} & \multicolumn{2}{c}{B3LYP} & \multicolumn{2}{c}{MP2} & &
\multicolumn{2}{c}{Exptl.$^c$} & \multicolumn{2}{c}{B3LYP} & \multicolumn{2}{c}{MP2} \\
    \hline
$\mu _a$         & 4&.0219~(50)  & 4&.119 & 3&.991 & & 3&.597~(59) & 4&.150 & 4&.021 & & 4&.1221~(31) & 4&.317 & 4&.151 \\
$\mu _{b/c}$$^d$ & 0&.6192~(267) & 0&.707 & 0&.702 & & 0&.984~(15) & 1&.095 & 1&.080 & & 0&.9026~(9)  & 0&.911 & 0&.915 \\
$\mu _{\rm tot}$ & 4&.0693~(275) & 4&.179 & 4&.052 & & 3&.729~(58) & 4&.292 & 4&.164 & & 4&.2197~(30) & 4&.413 & 4&.251 \\
    \hline
  \end{tabular}\\[2pt]
}
$^a$\footnotesize{See section~\ref{B3LYP} for details on quantum chemical calculations. 
     Numbers in parentheses are one standard deviation in units of the least significant figures.}\\
$^b$\footnotesize{Ref.~\cite{n-PrCN_dip}.}\\
$^c$\footnotesize{Ref.~\cite{c-PrCN_rot_2008}.}\\
$^d$\footnotesize{$c$-Component for $iso$- and $cyclo$-propyl cyanide; $b$-component 
     for the $anti$-conformer of $normal$-propyl cyanide.}
\end{table*}

Assuming the main principal axis of a nuclear quadrupole tensor to be aligned with 
a respective bond can lead to large errors in the derived principal values. 
An extreme case has been resolved by diagonalizing the $^{35}$Cl quadrupole tensor 
of SOCl$_2$ free of assumptions~\cite{SOCl2_rot_1994}. 
The value of $\chi _z$ was smaller by about one third and more compared to previous 
MW studies and very little $\pi$-bonding had to be inferred, in line with expectations, 
and with nuclear quadrupole resonance measurements, but in contrast to previous microwave results. 
A similar effect may occur if the main principal axis of the quadrupole tensor is 
in fact aligned with a respective bond, but the angles of this bond to the inertial axes 
are not known sufficiently well. 
There does not seem to be any electron diffraction study available for $iso$-propyl 
cyanide. Hence, the only experimental structural data available seem to be those 
from the more recent previous microwave study~\cite{i-PrCN_rot_1974}. 
Not knowing how reasonable these are, it was decided to perform 
quantum chemical calculations for further insight into the quadrupole coupling 
of $iso$-propyl cyanide and two related molecules. 
The results have been summarized in Table~\ref{Quadrupole_values}.


\begin{table*}
  \caption{Experimental $^{14}$N nuclear quadrupole coupling parameters $\chi _{ii}$ (MHz) 
    of $iso$-propyl cyanide estimated$^a$ off-diagonal coupling term $\chi _{ij}$ (MHz), 
    angle (deg) between quadrupolar $z$-axis and the CN bond, 
    and derived principal coupling value $\chi _z$ (MHz) and $\eta^b$ (unitless) 
    in comparison to values from quantum-chemical calculations$^c$ and corresponding values 
    for the $anti$-conformer of $normal$-propyl cyanide and for $cyclo$-propyl cyanide}
  \label{Quadrupole_values}
{\footnotesize
  \begin{tabular}{lr@{}lr@{}lr@{}lr@{}llr@{}lr@{}lr@{}lr@{}llr@{}lr@{}lr@{}lr@{}l}
  \hline
Parameter & \multicolumn{8}{c}{$i$-Propyl cyanide} & & \multicolumn{8}{c}{$a$-$n$-Propyl 
  cyanide} & & \multicolumn{8}{c}{$c$-Propyl cyanide} \\
\cline{2-9} \cline{11-18} \cline{20-27} 
 & \multicolumn{2}{c}{Exptl.} & \multicolumn{2}{c}{B3LYP} & \multicolumn{2}{c}{MP2} & \multicolumn{2}{c}{MP3} & &
\multicolumn{2}{c}{Exptl.$^d$} & \multicolumn{2}{c}{B3LYP} & \multicolumn{2}{c}{MP2} & \multicolumn{2}{c}{MP3} & &
\multicolumn{2}{c}{Exptl.$^e$} & \multicolumn{2}{c}{B3LYP} & \multicolumn{2}{c}{MP2}  & \multicolumn{2}{c}{MP3}\\
    \hline
$\chi _{aa}$        & $-$3&.938 & $-$4&.380 & $-$3&.584 & $-$3&.969 & & $-$3&.440 &
 $-$3&.843 & $-$3&.117 & $-$3&.476 & & $-$3&.460 & $-$3&.867 & $-$3&.123 & $-$3&.473 \\
$\chi _{bb}$        &    2&.111 &    2&.338 &    1&.919 &    2&.120 & &    1&.385 &
    1&.560 &    1&.227 &    1&.385 & &    1&.746 &    1&.895 &    1&.643 &    1&.717 \\
$\chi _{cc}$        &    1&.827 &    2&.042 &    1&.665 &    1&.849 & &    2&.056 &
    2&.283 &    1&.890 &    2&.092 & &    1&.714 &    1&.972 &    1&.480 &    1&.756 \\
$\chi _{ax}$$^f$    & $-$1&.278 & $-$1&.416 & $-$1&.218 & $-$1&.338 & & $-$2&.057 &
 $-$2&.285 & $-$1&.895 & $-$2&.103 & & $-$2&.013 & $-$2&.186 & $-$1&.801 & $-$2&.041 \\
$\angle (z,{\rm CN})$&    &$-$  &    0&.16  &    0&.27  &     &$-$  & &     &$-$  &
    0&.24  &    0&.39    &   &$-$    & &   &$-$  & $-$0&.04  &    0&.16  &     &$-$  \\
$\chi _{z}$         & $-$4&.209 & $-$4&.679 & $-$3&.853 & $-$4&.262 & & $-$4&.198 &
 $-$4&.679 & $-$3&.827 & $-$4&.260 & & $-$4&.151 & $-$4&.595 & $-$3&.744  & $-$4&.175  \\
$\eta$              &    0&.003 & $-$0&.001 & $-$0&.004 & $-$0&.001 & & $-$0&.021 &
 $-$0&.024 & $-$0&.012 & $-$0&.018 & & $-$0&.170 & $-$0&.175 & $-$0&.122 & $-$0&.178 \\
    \hline
  \end{tabular}\\[2pt]
}
$^a$\footnotesize{The B3LYP value was scaled with the $\chi _{kk}$ ratio of the experimental 
     and B3LYP value; see section~\ref{Discussion} for details.}\\
$^b$\footnotesize{$\eta = (\chi _x - \chi _y)/\chi _z$.}\\
$^c$\footnotesize{See section~\ref{B3LYP} for details on quantum chemical calculations. 
     Numbers in parentheses are one standard deviation in units of the least significant figures.}\\
$^d$\footnotesize{Ref.~\cite{n-PrCN_rot_1988}; uncertainties of the experimental $\chi _{ii}$ are in the last digit.}\\
$^e$\footnotesize{Ref.~\cite{c-PrCN_rot_2008}; uncertainties of the experimental $\chi _{ii}$ are in the last digit.}\\
$^f$\footnotesize{$x = c$ for $iso$- and $cyclo$-propyl cyanide; $x = b$ for the $anti$-conformer of 
      $normal$-propyl cyanide.}
\end{table*}

The quadrupolar $z$-axis is tilted away from the CN bond by very small amounts for all 
three molecules as one would have expected. There are (at least) three options for 
diagonalizing the $^{14}$N nuclear quadrupole tensor: first, one could start from the 
calculated $\theta _{za}$ angle, second, one could start from the calculated 
$\eta = (\chi _x - \chi _y)/\chi _z$ value, and finally, one could scale 
the calculated $\chi _{ac}$ value with the ratio of the observed and calculated 
$\chi _{bb}$ values; in the case of $n$-propyl cyanide, it would be $\chi _{ab}$ 
and $\chi _{cc}$, respectively. 
Ideally, each option should lead to the same result. In practice, the B3LYP values 
lead to slightly different estimates for the experimental $\chi _{ac}$ value: 
$-$1.271, $-$1.317, and $-$1.278, respectively. The magnitudes of these values are close 
to the 1.06~MHz determinable from the fit, see section~\ref{obs_spectrum}; but since 
the uncertainty of the latter is larger than its value, this agreement is not very meaningful. 
The values for $\eta$ differed slightly more, those for $\chi _z$ differed slightly less. 
The latter option was chosen since it resulted in magnitudes for the non-zero off-diagonal 
quadrupole coupling constant and the $\chi _z$ value that were between the other two. 
Trial fits employing the $cyclo$-propyl cyanide data set \cite{c-PrCN_rot_2008} revealed 
that $\chi _{ac} \approx 2.8$ can be obtained, albeit barely with significance if the stated 
uncertainties of 2~kHz for the FTMW data are used. These uncertainties seem conservative given an 
achievable rms value of 0.64~kHz and a data set of as many as 48 different FTMW lines.

Nuclear quadrupole coupling parameters calculated at the MP3/aug-cc-pwCVTZ level
were much closer to experimental values for all three molecules than values calculated at 
the MP2 or B3LYP levels. This may be seen as an indication that vibrational effects 
on the quadrupole parameters are possibly rather small for all these molecules. 
However, the agreement may as well be fortuitous as the level may be still too low. 
These parameters were calculated at the structural parameters obtimized at 
the MP2/aug-cc-pVTZ level because of computational time constraints, not at a 
structure optimized at the MP3 level. This will only have a sufficiently small effect 
if the structural parameters calculated at the MP3 level are sufficiently close 
to those at the MP2 level.

It is noteworthy that Durig and Li estimated values $\chi _{ac} = -1.22$~MHz and 
$\chi _z = -4.15$~MHz by assuming the $z$-axis to be aligned with the CN bond. 
These values are remarkably close to the present values in Table~\ref{Quadrupole_values}, 
suggesting that their preferred structural parameters are reasonable. 
They also deduced $\eta \approx 0$. 

The $\chi _z$ values, reported in Table~\ref{Quadrupole_values}, are also very similar 
among the three related molecules $iso$- and $cyclo$-propyl cyanide and 
the $anti$-conformer of $normal$-propyl cyanide 
in Table~\ref{Quadrupole_values}, and the small differences with respect to 
$cyclo$-propyl cyanide are probably significant. In fact, values very close to 
$-$4.2~MHz have been determined for a variety of organic cyanides: $\sim -$4.14~MHz 
for ethyl cyanide \cite{EtCN_eQq_1969}, $-$4.22~MHz for methyl cyanide 
\cite{MeCN_rot_2006,MeCN_rot_2009}, $\sim -$4.27 or $\sim -$4.18~MHz for 
vinyl cyanide \cite{VyCN_Sutter_1985,VyCN_Colmont_1997}, $-$4.237 for phenyl cyanide 
\cite{PhCN_rot_2008}, and even values for HC$_3$N and DC$_3$N, $-$4.3192 and 
$-$4.318~MHz, respectively \cite{HC3N-hfs_dip,DC3N_rot_2008}, are not very different. 
The situation for cyanoketene, $\chi _z \approx -$3.94~MHz \cite{cyanoketene_rot_2004}, 
is uncertain not only because of the large error bar, but also because small deviations 
from the assumed position of the $z$-axis may have a substantial effect on the 
$\chi _x$ and $\chi _z$ values because the CN bond is bent far away from all axes. 
Cyanocyclopropenylidene, in which the $z$-axis essentially agrees with the $a$-axis, 
and thus $\chi _z = -4.495$~MHz \cite{ccp_rot_1999}, and hydrogen cyanide, 
$\chi _z = -4.7078$~MHz \cite{HCN-hfs_dip}, have larger values. Similarly, 
a value of $-$4.61~MHz has been infered for acetyl cyanide aided by quantum 
chemical calculations \cite{AcCN_rot_2010}.

The value of $\eta$ is a measure of the deviation of the principal nuclear quadrupole tensor 
from cylindrical symmetry. If the nucleus belongs to an atom that is located in 
a terminal position of a formal single bond, $\eta$ is interpreted as a measure of 
the $\pi$-character, and hence the double bond character of that single bond. 
The CN bond, however, is probably as close to a true triple bond as is possible 
between two main group elements. Hence, $\eta$ is certainly not a direct measure 
of the $\pi$-character of that bond. It can, instead, be viewed as an indication, 
how much the triple bond character has been decreased. 
The magnitude of $\eta$ should be very small for a primary alkyl cyanide and increase 
somewhat for secondary and tertiary alkyl cyanides. Further increase should occur 
if the cyano group is attached to a small saturated ring molecule, and even more so, 
if it is attached to an atom which is involved in $\pi$-bonding. As can be seen in 
Table~\ref{Quadrupole_values}, $\eta$ is close to zero for the $anti$-conformer of 
$normal$-propyl cyanide. However, it is much closer to zero for $iso$-propyl cyanide, 
and it is zero because of symmetry in $tertiary$-butyl cyanide. The nuclear quadrupole 
tensor in its principal axis system deviates considerably for $cyclo$-propyl cyanide, 
and it is not so common that a diagonal quadrupole tensor in its inertial axis system 
is more symmetric than in its principal axis system. On the other hand, 
$\eta$ is quite close to zero in vinyl cyanide \cite{VyCN_Sutter_1985,VyCN_Colmont_1997}, 
suggesting that there is little interaction between the $\pi$-electrons of the 
vinyl group and those of the cyano group. Together with the very similar values for 
$\chi _z$, this suggests that the CN group in all these molecules has a triple bond. 
Thus, it cannot be surprising that for all three investigated conformers of $n$-butyl 
cyanide good agreement with experimentally measured quadrupole coupling parameters 
was obtained when the cylindrically symmetric quadrupole tensor of methyl cyanide was 
rotated into the corresponding inertial axis systems~\cite{n-BuCN_rot_1997}
This bonding model appears to be a good approximation even for phenyl cyanide, 
since $\eta = -0.080$ \cite{PhCN_rot_2008} is also fairly close to zero. 
The principal quadrupole tensors of acetyl cyanide, cyanoketene, and 
cyanocyclopropenylidene display larger deviations from cylindrical symmetry 
with $\eta$ values of about 0.19 \cite{AcCN_rot_2010}, $-$0.37 
\cite{cyanoketene_rot_2004}, and 0.38 \cite{ccp_rot_1999}, respectively. 
As mentioned above, the value of cyanoketene is somewhat more uncertain. 
A factor of 1/4 for $\chi _{bb} - \chi _{cc}$ of cyanopropenylidene has been forgotten 
in the publication~\cite{ccp_rot_1999}. Finally, a value of $\eta \approx 0.2$ 
was found as typical for cyano groups attached to carbonyl groups \cite{AcCN_rot_2010}, 
based on several examples. 

It may, at first sight, be surprising that nuclear magnetic spin-rotation parameters 
can be determined with some accuracy for a nucleus having a small magnetic $g$-factor 
and being in a molecule with comparatively small rotational constants. 
Comparison between experimental and theoretically calculated values in 
Table~\ref{parameters} are favorable given the available accuracies. 
$^{14}$N spin-rotation parameters with very similar values can be determined for the related 
$cyclo$-propyl cyanide employing the published data set~\cite{c-PrCN_rot_2008}. 
Such parameters have been determined from FTMW spectroscopy and published, e.g., 
for the slightly lighter vinyl cyanide~\cite{13C-EtCN-det_Sgr_B2,VyCN_Colmont_1997} 
or the fairly heavy SOCl$_2$~\cite{SOCl2_rot_1994}.

As can be seen in Table~\ref{Dipole_values}, a very accurate $\mu _a$ value of 4.0219~(50)~D 
has been determined for $iso$-propyl cyanide, its accuracy is mainly limited by the 
calibration uncertainties as mentioned in section~\ref{Dipole}. The previously determined 
value of 4.05~(2)~D \cite{i-PrCN_rot_1974} is essentially the same, 
only its uncertainty is somewhat larger. 
The $c$-component was determined here as 0.6192~(267)~D. The uncertainty is much larger 
than that of the $a$-component because all Stark-shifts are dominated by the latter component. 
This value does not agree at all with 1.4~(2)~D determined in an earlier study 
\cite{i-PrCN_rot_1974}. This large difference results in a factor of 5 weaker line 
intensities. In fact, the average line intensity ratio between $a$ and $c$-type transitions 
is 42, about a factor of two more than was estimated in Ref.~\cite{i-PrCN_rot_1967}.

The experimental dipole moment components of $iso$- as well as $cyclo$-propyl cyanide 
agree rather well with those from quantum-chemical calculations which are all presented 
in Table~\ref{Dipole_values}. This is not the case for the $anti$-conformer of 
$n$-propyl cyanide. Assuming deviations to be similar to those observed 
for the other two molecules, one would expect $\mu _a \approx 4.0$~D instead of 
$\sim$3.6~D from experiment \cite{n-PrCN_dip}. It may be warranted to revisit 
the Stark effect measurements of $n$-propyl cyanide. The experimental and theoretical 
values for $\mu _b$, on the other hand, are in accord.
It may be noted that in $iso$-propyl cyanide the angle of the dipole vector with 
the $a$-axis is quite similar to the angle between the CN bond and the $a$-axis, 
the latter is only 2.3$^{\rm o}$ bent farther away, but this is an exception. 
For both $cyclo$-propyl cyanide as well as the $anti$-conformer of $n$-propyl cyanide, 
the CN bond is bent farther away from the $a$-axis as well as from the dipole vector.

Irrespective of the controversy concerning the size of the $b$-dipole moment component 
of ethyl cyanide \cite{EtCN_600GHz_mu_b_2010}, its total dipole moment is about 4.05~D, 
marginally larger than 3.922~D of methyl cyanide \cite{MeCN-dipole}, and also close 
to $\sim$3.9~D of vinyl cyanide \cite{VyCN_Sutter_1985}, showing little variation 
among monocyanides of saturated or almost saturated hydrocarbons.


\section{Conclusions}
\label{Conclusions}

The rotational spectrum of $iso$-propyl cyanide has been investigated from the MW 
well into the sub-mm region at 600~GHz. Accurate rotational constants as well as 
centrifugal distortion parameters have been obtained. These are sufficiently accurate 
to search for the molecule in space in all regions relevant for radio-astronomy. 
The isomer $n$-propyl cyanide was detected in the prolific Galactic center 
hot-core source Sgr~B2(N) in the course of a line survey which covered 
the entire 3~mm region along with sections at 2 and 1.3~mm \cite{n-PrCN_EtFo-det}. 
The derived rotational temperature was 150~K, and is fairly common not only for 
saturated cyanides, but also for the warmer parts of such hot cores. 
The attempts to identify $iso$-propyl cyanide in that molecular line survey of 
Sgr~B2(N) were not conclusive and will be discussed with searches for other, 
related molecules in detail elsewhere. Further searches, in other sources or 
other frequency regions, can now be performed. 
The Boltzmann peak in the rotational spectrum of $iso$-propyl cyanide at 150~K 
occurs at 230~GHz. The strongest transitions in the 3~mm region are only about 
a factor of three weaker than those around the Boltzmann peak. 
Modeling of the data obtained in that line survey suggest that molecules 
as large as aminoacetonitrile \cite{AAN-det}, ethyl formate, or 
$n$-propyl cyanide \cite{n-PrCN_EtFo-det} and with similar line intensities 
can be detected best in or near the 3~mm region in the case of observations 
with a single dish radio-telescope and if only the rotational spectrum is considered. 
The single most important reason for this limitation is the plethora 
of spectral features caused by many small to medium sized molecules 
with fairly high abundances. 
Molecules with stronger lines, because of larger abundance, dipole moment, 
or rotational constants, may be detectable more favorably at frequencies 
closer to or even beyond the Boltzmann peak. For searches for molecules with weaker lines, 
lower frequencies are recommended. Resorting to array instruments, such as the 
Atacama Large Millimeter Array (ALMA) may well reduce the problem of line confusion, 
but the exact extent remains to be explored. At the upper frequency end of the present 
investigations, the intensities of the strongest lines at 150~K have dropped 
already by a factor of 40 from those at the Boltzmann peak, 
suggesting that searches for this molecule at such high frequencies are not reasonable. 
But even if ALMA should reveal even denser and hotter parts of the ISM 
in which molecules such as $iso$-propyl cyanide or complexer ones may be detected, 
it should be kept in mind that extrapolations of the rotational spectrum should be 
reasonable up to about twice the upper frequency, which is well beyond the final ALMA range.

Besides rotational and centrifugal distortion parameters, greatly improved $^{14}$N 
nuclear quadrupole coupling parameters as well as approximate values for the 
nuclear magnetic spin-rotation parameters have been obtained. Aided by quantum-chemical 
calculations, it was shown that the quadrupole tensor in its principal axis system 
is essentially identical to that of, e.g., methyl cyanide, as one would expect and 
as has been assumed previously. 

An earlier value for the dipole moment component along the $a$-axis was confirmed 
and improved, and the value for the $c$-axis was revised downward considerably.



\section*{Acknowledgements}

We appreciate the help of Frank Lewen and Christian P. Endres 
with the spectrometer systems. 
H.S.P.M. is very grateful to the Bundesministerium f\"ur Bildung und 
Forschung (BMBF) for financial support aimed at maintaining the 
Cologne Database for Molecular Spectroscopy, CDMS. This support has been 
administered by the Deutsches Zentrum f\"ur Luft- und Raumfahrt (DLR). 
A.C. and A.W. thank the French National Program PCMI (CNRS/INSU) and the
Observatoire Midi-Pyr\'en\'ees for funding. J.-U.G. acknowledges support 
by the Deutsche Forschungsgemeinschaft (DFG) and the Land Niedersachsen.



\bibliographystyle{elsarticle-num}
\bibliography{<your-bib-database>}




\end{document}